\documentclass[aip,graphicx]{revtex4-1}
\usepackage{graphicx}
\usepackage{color}
\usepackage{textcomp}
\usepackage{amssymb}
\usepackage{epstopdf}

\draft 

\begin{document}


\title{Apparatus for simultaneous DLS-SANS investigations of dynamics and structure in soft matter} 



\author{V. Nigro}
\thanks{Present address: ENEA C.R. Frascati, FSN-TECFIS-MNF Photonics Micro and Nanostructures Laboratory, 00044 Frascati, Roma, Italy}
\author{R. Angelini}
\email{roberta.angelini@cnr.it}
\affiliation{Institute for Complex Systems, National Research Council (CNR-ISC) and Department of
Physics, Sapienza University of Rome, Piazzale A. Moro 2, 00185, Rome, Italy}
\author{S. King} 
\affiliation{ISIS Pulsed Neutron \& Muon Source, STFC Rutherford Appleton Laboratory, Harwell Campus, Didcot, Oxon OX11 0QX, United Kingdom}
\author{S. Franco}
\author{E. Buratti}
\author{F. Bomboi}
\affiliation{Institute for Complex Systems, National Research Council (CNR-ISC) and Department of Physics, Sapienza University of Rome, Piazzale A. Moro 2, 00185, Rome, Italy}
\author{N. Mahmoudi}
\affiliation{ISIS Pulsed Neutron \& Muon Source, STFC Rutherford Appleton Laboratory, Harwell Campus, Didcot, Oxon OX11 0QX, United Kingdom}
\author{F. Corvasce}
\affiliation{Institute of Structure of Matter, National Research Council (CNR-ISM) Via del Fosso del Cavaliere 100, I-00133 Roma, Italy}
\author{R. Scaccia}
\affiliation{Institute of Structure of Matter, National Research Council (CNR-ISM) Via del Fosso del Cavaliere 100, I-00133 Roma, Italy}
\author{A. Church}
\affiliation{ISIS Pulsed Neutron \& Muon Source, STFC Rutherford Appleton Laboratory, Harwell Campus, Didcot, Oxon OX11 0QX, United Kingdom}
\author{T. Charleston}
\affiliation{ISIS Pulsed Neutron \& Muon Source, STFC Rutherford Appleton Laboratory, Harwell Campus, Didcot, Oxon OX11 0QX, United Kingdom}
\author{B. Ruzicka}
\email{barbara.ruzicka@cnr.it}
\affiliation{Institute for Complex Systems, National Research Council (CNR-ISC) and Department of Physics, Sapienza University of Rome, Piazzale A. Moro 2, 00185, Rome, Italy}

\date{\today}
\begin{abstract}
Dynamic Light Scattering (DLS) and Small-Angle Neutron Scattering (SANS) are two key tools with which to probe the dynamic and static structure factor, respectively, in soft matter. Usually DLS and SANS measurements are performed separately, in different laboratories, on different samples and at different times. However, this methodology has particular disadvantages for a large variety of soft materials which exhibit high sensitivity to small changes in fundamental parameters such as waiting times, concentration, pH, ionic strength, etc. 
Here we report on a new portable DLS-SANS apparatus that allows one to simultaneously measure both the microscopic dynamics (through DLS) and the static structure (through SANS) on the same sample. The apparatus has been constructed as a collaboration between two laboratories, each an expert in one of the scattering methods, and was commissioned on the \textit{LOQ} and \textit{ZOOM} SANS instruments at the ISIS Pulsed Neutron \& Muon Source, U.K.
\end{abstract}


\maketitle 

\section{\label{sec:level1}INTRODUCTION}

Soft materials are ubiquitous in a variety of industries including foods, pharmaceuticals, personal care products, and cosmetics. They exhibit unique behaviours that are often fundamental to their applications in those industries \cite{NievesWiley2016} but which stem from their microscopic properties \cite{Springer2008, LikosPhysRep2001}. Soft materials also offer the possibility to investigate many of the key topics in soft condensed matter in general, such as slow reactions and kinetics, ageing processes, gelation, controlled release, food processing, micellar growth, and aggregation phenomena. Correlating the microscopic behaviour with these macroscopic processes is thus a fertile ground for research.

Light scattering has long been used for the characterization of soft matter and dynamic light scattering (DLS) has demonstrated itself to be a crucial and indispensable technique with which to investigate the dynamics \cite{PuseyScattering2002}. DLS measures the stochastic temporal
variations in scattered laser light, resulting in a time autocorrelation function describing the timescales of mutual diffusive motion of the scattering objects. This information can then be used to derive the hydrodynamic size of objects in suspension \cite{KaszubaJNR2008, DalgleishFRI1995}. However, the autocorrelation function itself is especially useful for investigating the microscopic dynamics of soft matter systems, including colloids \cite{PuseyNat1986, PhamScience2002, EckertPRL2002}, polymers \cite{Burchard1989}, microgels \cite{NigroJCIS2019, NigroMacromol2020} proteins \cite{YuJPS2013}, and vesicles \cite{GrimaldiFrontiers2019}, and for characterising phenomena like sol-gel transitions \cite{KroonPRE1996} or the formation of arrested states through changing concentration or waiting time (ageing) \cite{PhilippePRE2018,  VanMegenPRL1993, RuzickaPRL2004}. All this is possible because the technique studies spatial perturbations of the order of a micron or less (a typical scattering vector is of the order of $Q$ = 10$^{-2}$ nm$^{-1}$), but also probes time scales in the range of nanoseconds and above. These two conditions typically characterise a wide range of soft matter systems.

Techniques such as (Ultra-)Small-Angle Scattering, whether performed with X-rays ((U)SAXS) or neutrons ((U)SANS) are capable of probing soft matter on comparable, to somewhat complementary length scales as light scattering, typically from a few nm to hundreds of nm \cite{Zemb2002, GrilloSpringer2008, Imae2011}. However, compared to DLS these techniques are only capable of studying the time-averaged structure in the system. The best laboratory SAXS or SANS instruments might achieve a time resolution of around 100 milliseconds in an optimal system (though sub-millisecond time resolution has been demonstrated with a rare form of SANS called TISANE \cite{GlinkaAppCryst2020}), whilst even Synchrotron SAXS would be hard-pressed to achieve better than 100 microseconds. On the other hand, the great advantage of neutrons is that they permit one to selectively highlight, or suppress, the scattering from one or more components in a complex soft matter system using deuteration. This 'contrast variation' approach is possible with light and X-rays (by varying refractive index or electron density, respectively), but in practical terms is extremely difficult to execute. There is, therefore, a clear scientific benefit to performing simultaneous DLS-SANS measurements on soft matter systems arising from the synergy between the two techniques. The clear disadvantage of SANS, of course, is that (at present) it can only be performed at Large-Scale Facilities.

Today, the use of multiple experimental techniques is recognised as being of fundamental importance to understanding the properties of real-world materials. In particular, combining information about both structural and dynamical behaviour is imperative since it is the complex interplay between structure and dynamics that determines the microscopic relaxation processes responsible for the macroscopic mechanical properties of a material. This knowledge is crucial for a detailed understanding of a samples properties and to facilitate the tailored design of materials with specific properties \cite{RuzickaNatMat2011, AngeliniNC2014}. Within this framework, reliably performing simultaneous DLS and SANS measurements would represent a breakthrough in helping develop our understanding of this underlying and often complex behaviour that simply cannot be uniquely determined through the use of a single technique.

Given the above, it will not surprise the reader to learn that combining DLS with SAXS/SANS has been a goal of several groups in recent years. Vavrin et al. \cite{VavrinRSI2007} augmented a high-pressure SANS sample environment with a DLS. Nawroth et al. \cite{NawrothMolPharm2011}, and later Schrader et al. \cite{SchraderCrystGrowth2018}, both combined DLS with Stop-Flow SANS. Schwamberger et al. \cite{SchwambergerNIMPR2015} implemented DLS-SAXS on a flow-through capillary for online metrology purposes whilst Falke et al.\cite{Falke_JSynchRad_2017} designed a multi-channel DLS system for the BioSAXS endstation at PETRA III. 
However, because the vast majority of soft matter SANS experiments are conducted on fluid samples in cuvettes which are both optically and neutronically transparent, delivering a capability for DLS-SANS with a sample changer has remained a key aspiration of the SANS community.

The aim of the present paper is to report on a new compact, portable, DLS apparatus, conceived, designed and implemented by a team from the Institute for Complex Systems (ISC), CNR, Italy, working in collaboration with SANS scientists and engineers from the ISIS Pulsed Neutron \& Muon Source, UK. The apparatus is fully-enclosed to meet stringent safety regulations, has a relatively small footprint, and includes a temperature-controlled sample changer. When installed on a SANS instrument the apparatus permits complementary and simultaneous information on the microscopic structure (through SANS) and dynamics (through DLS) of soft matter systems to be acquired. In this paper we validate the performance of the instrument by performing simultaneous DLS and SANS measurements on dilute polymer nanoparticle samples and compare the radii derived through the two techniques.

\section{DESIGN CRITERIA}
Whilst SAXS/SANS instruments are often highly adaptable to the sample environment needs of an experiment, equipment complexity and experimental turnaround time are generally divergent quantities. This becomes an important consideration at Large-Scale Facilities where the inherent cost of beam time is high and there is an ethos of minimizing downtime.  Laser safety is another important consideration in these multi-user working environments. Whilst laser hazards can always be mitigated by engineered controls and interlocks, if these are infrastructural, as distinct from local to the apparatus, then the portability of the apparatus is sacrificed.

With these considerations in mind, the new apparatus has been designed to satisfy several competing requirements: an enclosed, ‘drop in’/’modular’ apparatus with a compact footprint, sufficiently robust to facilitate use on any of the four SANS instruments at ISIS (or in principle elsewhere), with its own interlock system, easy to align (both neutronically and optically) in a reproducible manner, and incorporating a temperature-controlled sample changer (to reduce operator interaction with the laser, to make more efficient the use of neutron beamtime, and indeed to facilitate useful science).

From the outset, one of the most challenging aspects of the design was the need to maintain as minimal an air path for the neutrons as possible. This was because the 'cold' neutrons used in SANS are significantly attenuated and scattered by the much more massive gas molecules in air. As evacuating the apparatus was not considered a practical solution, given most soft matter samples are fluids, this meant that the feasibility of the design depended on one important technological challenge: namely could the optical elements of the apparatus be  miniaturized sufficiently, whilst retaining good quality optical data. Achieving this would naturally also help to make the apparatus compact and portable.

To find out, a simple prototype DLS apparatus was constructed at the CNR-ISC in order to experiment with different optical geometries and beam paths. The efficiency of these setups were then characterized in terms of the quality of the autocorrelation function acquired in a given time from a silica nanoparticle dispersion standard, using the same sort of sample cuvette and sample volume ($\approx$250 microlitre) as would be used for a SANS experiment.

\begin{figure}[ht]
\centering
\includegraphics[trim=4cm 4cm 1cm 2cm,clip, width=1.1\textwidth]{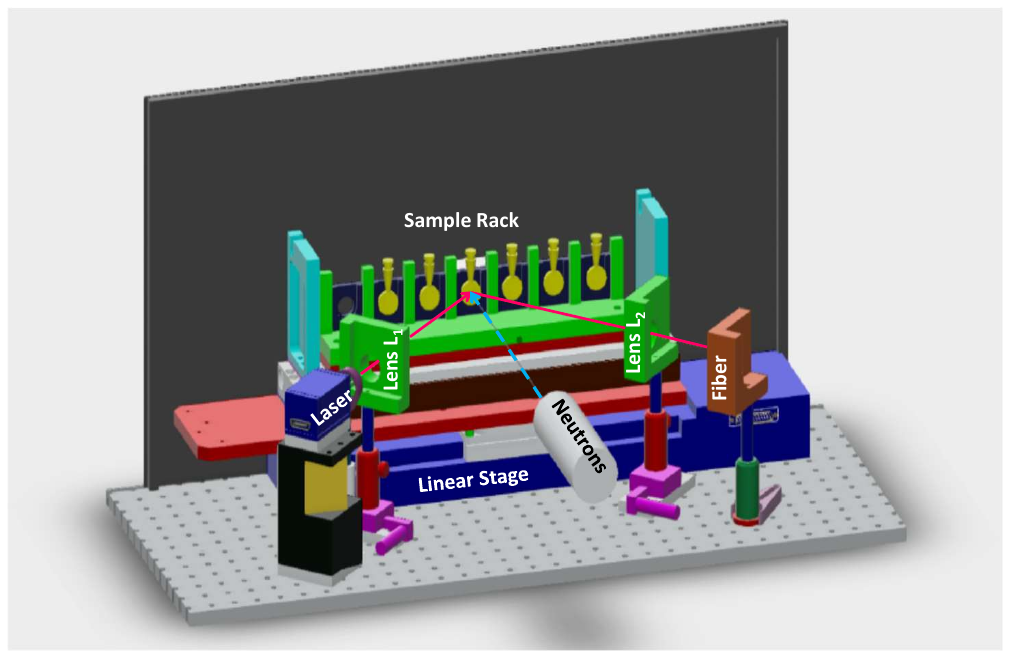}
\caption{3D technical drawing of modular DLS apparatus. The enclosing cover, with neutron transparent beam windows, is not shown} \label{f1}
\end{figure}

From combining the design criteria with the prototyping we then arrived at the final design concept illustrated in Fig.~\ref{f1}. A 'lift-on/lift-off', interlocked, enclosure, presenting a robust physical boundary capable of adequately containing the laser radiation generated within it, protects operators on the outside, and helps to exclude dust and facilitate temperature control.

\section{APPARATUS DESCRIPTION}
\begin{figure}[ht]
\centering
\includegraphics[trim=2cm 5cm 1cm 3cm,clip, width=1.1\textwidth]{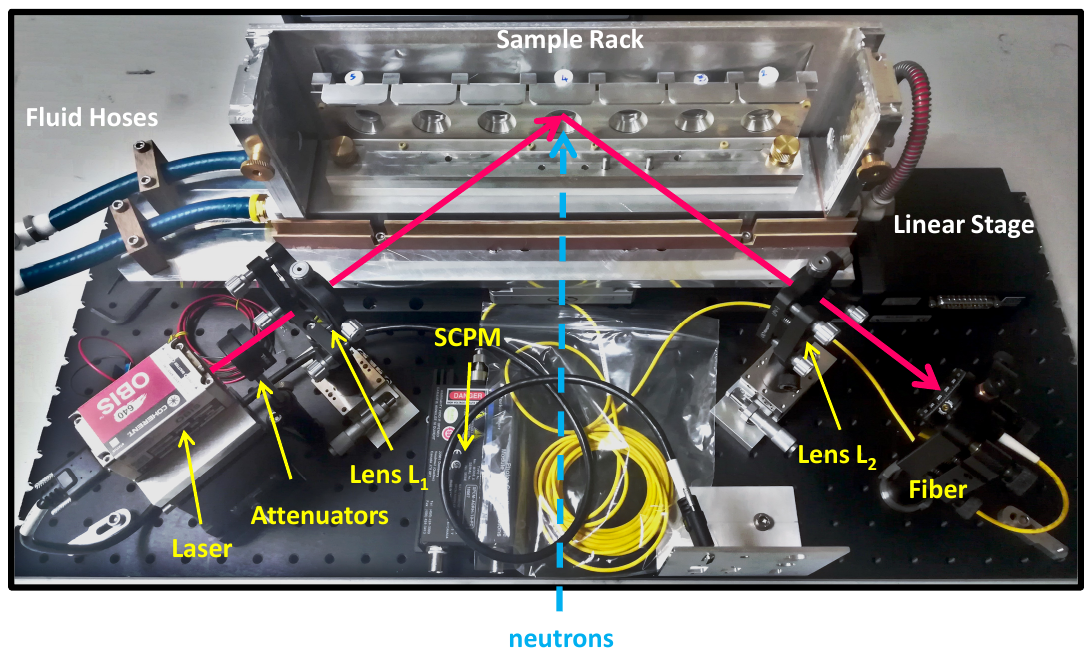}
\caption{Photograph of the DLS-SANS apparatus without the enclosure. The cover of the sample rack assembly, which aids temperature control, is also removed for clarity. The red arrows indicate the path of the laser beam, while the dashed cyan arrow indicates the path of the neutron beam.} \label{f2}
\end{figure}

The DLS-SANS apparatus (see Fig.~\ref{f2}) comprises a black anodized aluminium optical breadboard (780 x 350 x 13 mm, Thorlabs) onto which all of the system is mounted. The output from a solid state laser of 100 mW power at $\lambda$ = 642 nm (Coherent\textsuperscript{\textregistered} OBIS$^{TM}$ 640 LX) is attenuated through an interchangeable optical filter and focused onto the center of a sample cuvette (1 or 2 mm pathlength, e.g. Hellma GmbH Type 120-QS or Starna Scientific Type 32-Q) through the lens L$_1$. A short linear translation stage (M-ILS250PP, Newport Instruments) permits movement of a temperature-controlled sample changer assembly on which is mounted a 7-position rack (in-house design, ISIS Pulsed Neutron \& Muon Source) that houses the sample cuvettes. Temperature control is achieved using an external circulating fluid bath and monitored with thermocouples on the sample rack.

The laser beam scattered from the sample is then collected by the lens L$_2$ and directed on to a single mode fiber feeding a Single Photon Counting Module (SPCM-AQRH-13-FC, PerkinElmer). The resulting electrical signal is then passed to a LSi logarithmic correlator \cite{LSI} that computes the intensity autocorrelation function.

The optical elements (Thorlabs) are mounted on micrometers that permit horizontal (H) and vertical (V) translations, tilt (T) and focus (F) movements to facilitate alignment. In particular, the following movements are allowed: laser (H, V), lens (H, V, T, F) and fiber (H, V, T). The scattering angle is $\theta_1$=105\textdegree  which, according to the relation $Q$ = $|\vec Q|$ = (4$\pi n/ \lambda$) sin($\theta$/2) where $n$ is the  index of refraction (note $n$=1 for neutrons), corresponds to a scattering vector $Q$ = 0.021 nm$^{-1}$ for aqueous samples. 

\begin{figure}
\centering
\includegraphics[trim=3cm 4cm 3.5cm 3cm,clip, width=1\textwidth]{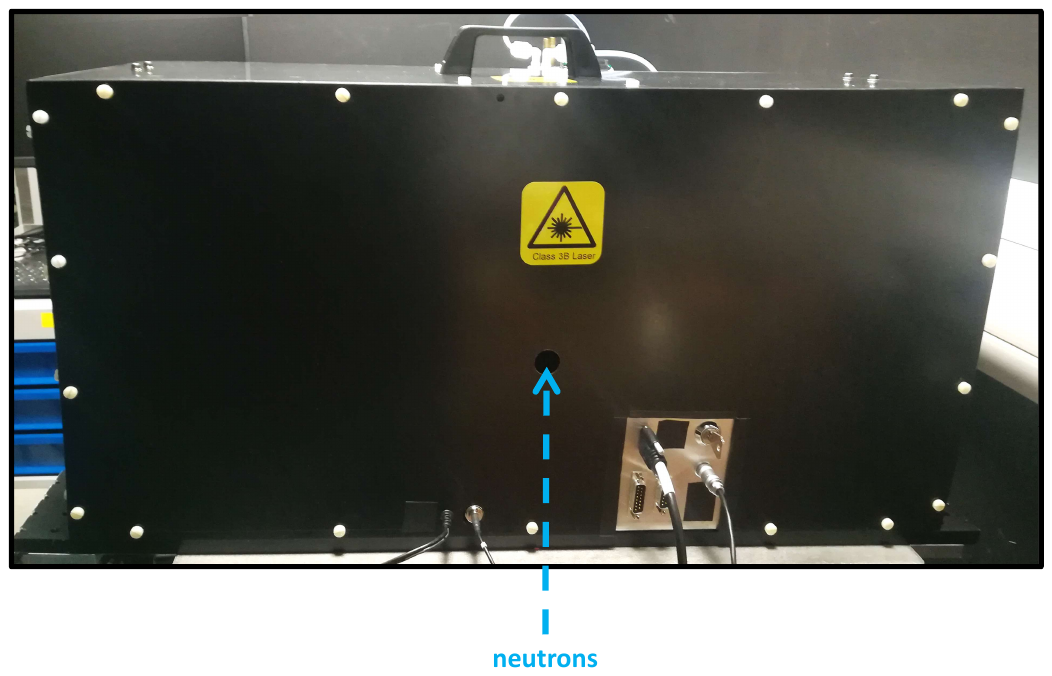}
\caption{Photograph of the DLS-SANS enclosure. The dashed cyan arrow indicates the path of the neutron beam.} \label{f3}
\end{figure}

A rigid black plastic enclosure (350 mm high), but with an aluminium back wall, fits over, and locates on, the breadboard, completely enclosing everything inside (see Fig.~\ref{f3}). The only apertures in the cover are for the passage of the neutron beam (and these are normally covered by aluminium foil to preserve light-tightness), a cut-out on the left side where the fluid hoses emanate from, and a cut-out on the front wall permitting access to a connector panel. Neither of the last two apertures exposes the direct or reflected laser beam. There are no optical windows on the enclosure. The reason for aluminium back wall, behind the samples, is that some neutrons will be scattered by the sample to large angles and ultimately pass through the back wall instead of through the foil-covered neutron beam exit window. If these were to subsequently reach the neutron detector they would contaminate the SANS from the sample. Neutronically, there is much less SANS from aluminium than from plastic. An alternative strategy would be to simply glue 1 mm thick cadmium sheet (a neutron absorber) onto the aluminium back wall.

The presence of the enclosure and its lid engages microswitches that form part of the laser interlock system. For sample changes, undoing 6 thumbscrews allows just the lid of the enclosure to be removed, but this operation breaks the laser interlock. For optical alignment of the apparatus, a key switch on the front connector panel (visible in Fig.~\ref{f3}) allows suitably authorized persons to operate the laser without the lid, or indeed the enclosure, in place.

The overall dimensions of the apparatus are 780 x 350 x 363 mm (length x width x height) and its total mass is about 30 Kg.

\section{EXPERIMENTAL METHODS}
\begin{figure}
\centering
\includegraphics[trim=2cm 8cm 1cm 1cm,clip, width=1.1\textwidth]{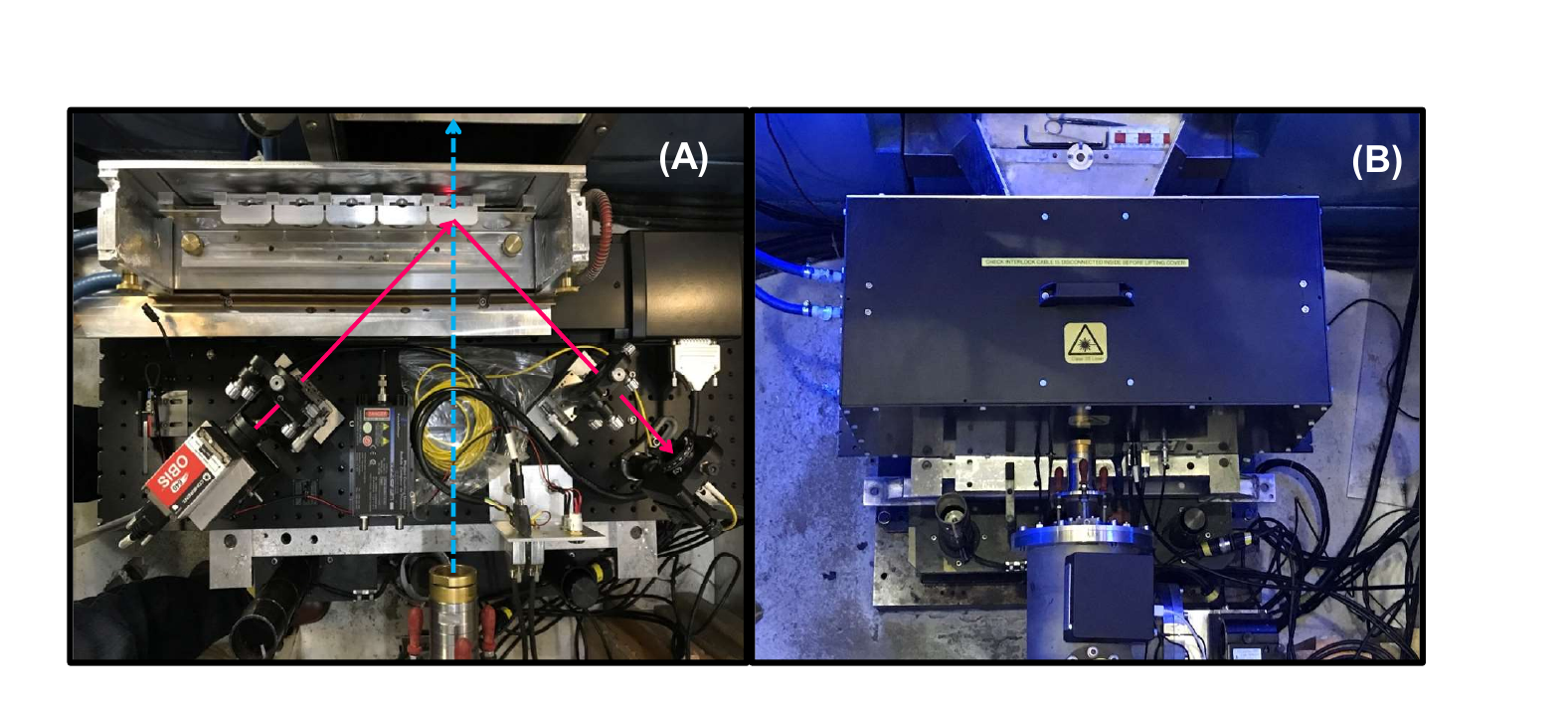}
\caption{Photographs of the DLS-SANS apparatus installed on the \textit{LOQ} SANS instrument at ISIS (A) without and (B) with the enclosure installed. The red arrows indicate the path of the laser beam, while the dashed cyan arrow indicates the path of the neutron beam.} \label{f4}
\end{figure}

\begin{figure}
\centering
\includegraphics[trim=1cm 8cm 1cm 1cm,clip, width=1.05\textwidth]{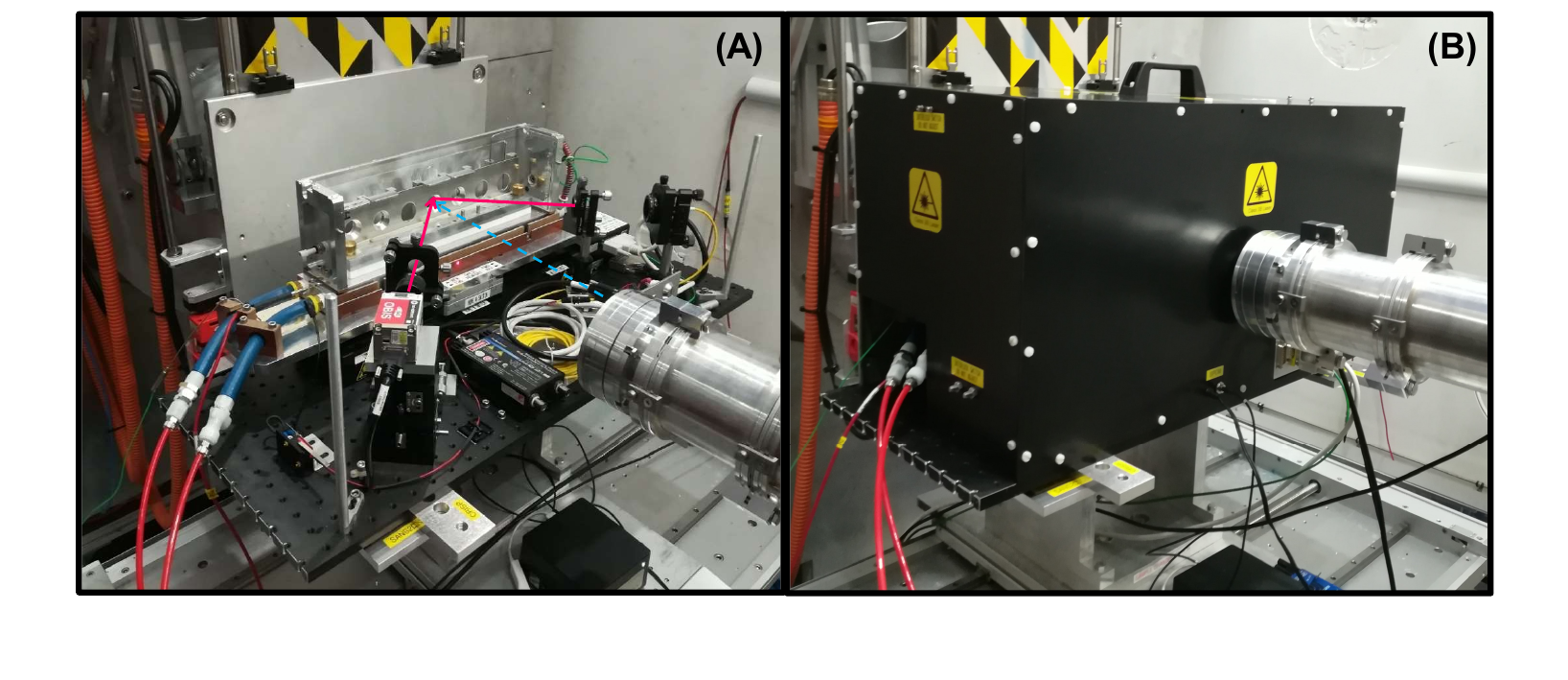}
\caption{Photographs of the DLS-SANS apparatus installed on the \textit{ZOOM} SANS instrument at ISIS (A) without and (B) with the enclosure installed. The red arrows indicate the path of the laser beam, while the dashed cyan arrow indicates the path of the neutron beam.} \label{f5}
\end{figure}

The apparatus has so far been used on both the \textit{LOQ} and \textit{ZOOM} SANS instruments at ISIS, with excellent performance. Photographs of the apparatus installed on the \textit{LOQ} instrument, with and without the enclosure, are shown in Fig.~\ref{f4}(A) and Fig.~\ref{f4}(B), respectively. Similarly, photographs of the apparatus installed on the \textit{ZOOM} instrument are shown in Fig.~\ref{f5}(A) and Fig.~\ref{f5}(B), respectively. Below we report DLS-SANS data obtained on the \textit{LOQ} instrument.

\subsection{SANS measurements}
SANS measurements were performed on the \textit{LOQ} instrument located on Target Station 1 at the ISIS Pulsed Neutron \& Muon Source (STFC Rutherford Appleton Laboratory, Didcot, U.K) \cite{HeenanJAC1997} This is a time-of-flight diffractometer with a typical time-averaged flux at the sample of ~3x10$^5$ cm$^{-2}$ s$^{-1}$ at 25 Hz, which utilizes a polychromatic incident neutron beam spanning wavelengths of 0.22 $\leq$ $\lambda_0$ $\leq$ 1.0 nm, simultaneously recorded on two, 2-dimensional 'area' detectors with overlapping angular coverage, to provide a very wide dynamic range in scattering vector of 0.07 $\leq$ $Q$ $\leq$ 14 nm$^{-1}$ but with fixed sample-detector distances of 0.5 and 4.1 m. This type of SANS instrument is ideal for studies where the length scales of interest are uncertain, or broad, or indeed evolving, as time-consuming instrumental re-configurations and re-calibrations are unnecessary. The neutron beam incident on the sample was collimated to 8 mm in diameter.

Using the Mantid framework (version 4.1.0) \cite{Mantid} each raw SANS dataset was corrected for the incident neutron wavelength distribution, the detector efficiency and spatial linearity, the measured sample transmission, and the sample path length, before being radially averaged and converted into the coherent elastic differential scattering cross-section ($\partial\Sigma/\partial\Omega (Q)$) as a function of $Q$, hereafter simply referred to as intensity, $I(Q)$. This process also merged the data from the two detectors onto a common $Q$ scale. The contribution from instrumental (vacuum windows, etc) and sample (cuvette, incoherent scattering, etc) background scattering was removed by subtracting the  SANS measured from a pure D$_2$O sample. Finally, the fully-reduced SANS data was placed on an absolute intensity scale by reference to the scattering from a partially-deuterated solid polystyrene blend standard sample of known molecular weight, measured with the same instrument configuration \cite{WignallJAC1987}. SANS data analysis was performed using the Sasview software \cite{Sasview}.

\subsection{DLS-SANS measurements}
To test the performance of the DLS-SANS apparatus, measurements were first made on a light scattering standard sample of known size, then subsequently on a colloidal suspension of microgel particles of contemporary interest.

One complicating factor worth mentioning is that because DLS requires samples of low turbidity, from a SANS perspective samples can be of much lower concentration than might normally be considered suitable, or even sensible. Since SANS is an inherently count rate limited technique anyway, the only recourse is to significantly increase counting times. In our tests we did explore some of this envelope. In general, we found that samples with a weight concentration C$_w$ $\le$0.3 \% worked best. The lowest concentration we measured was a microgel sample at C$_w$=0.03 \%, but obviously the size of the particles being studied, and their refractive index, will have had a bearing on this. The \textit{LOQ} instrument is, by modern standards, a relatively low flux instrument. On the \textit{ZOOM} instrument we were able to effectively halve counting times.

\subsubsection{Standard Sample}

An aqueous polystyrene latex standard (PLS; NanosphereTM 3000 Series, Fisher Scientific UK) was diluted with D$_2$O from 1\% solids to a weight concentration of C$_w$=0.06 \% . The latex spheres were certified as having a nominal radius of 31 $\pm$ 3 nm. 

\begin{figure}
\centering
\includegraphics[trim=0cm 20cm 0cm 0cm,clip, width=.99\textwidth]{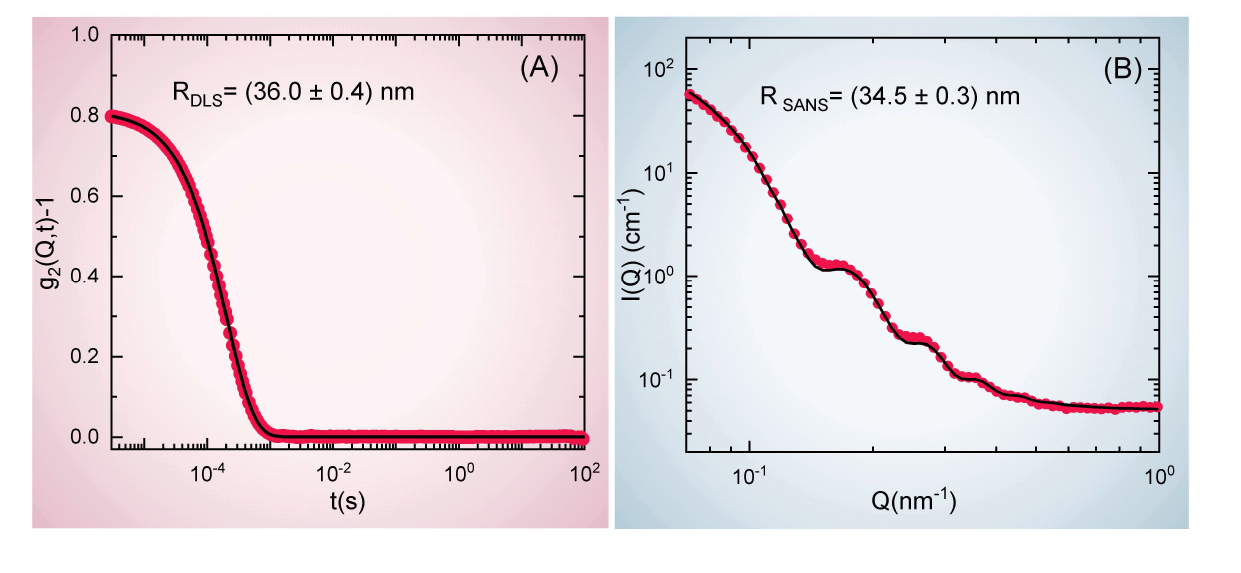}
\caption{(A) Intensity autocorrelation function and (B) SANS intensity as simultaneously obtained from the PLS sample using the DLS-SANS apparatus. The continuous black lines superimposed on the data (symbols) are fits to Eq.\ref{eq1} and Eq.\ref{eq3}, respectively.} \label{f6}
\end{figure}

Representative examples of the resulting DLS intensity autocorrelation functions, $g_2(Q,t)-1$, are shown in Fig.~\ref{f6}(a) and of the SANS data, $I(Q)$, in Fig.~\ref{f6}(b). The SANS data were acquired over ~6.5 hours. 

The particle size was then extracted from the intensity autocorrelation function by fitting it to the conventional Kohlrausch-William-Watts (KWW) expression \cite{KohlrauschAnnPhys1854, Williams1970, NigroSM2017}: 

\begin{equation}
g_2(Q, t) =1 + b[exp(-t/\tau)^{\beta}]^2
\label{eq1}
\end{equation}

where $\tau$  is the relaxation time and $\beta$ describes the deviation from the simple exponential decay ($\beta$ = 1) which gives a measure of the distribution of relaxation times. In this case a value of $\beta \cong$ 1 was found as was expected given the low size polydispersity of the standard sample. Alternatives to Eq.\ref{eq1}, such as Cumulant analysis or the Contin algotithm, can be successfully employed for the determination of the average particle size and width of the particle size distribution, even if some limitations may result from a lack of robustness when dealing with more complex systems.

The  diffusion coefficient $D$ was then derived from $\tau$ using the relation $\tau=1/DQ^2$, and from that the hydrodynamic radius of the particles was estimated using the Stokes-Einstein relationship for spherical particles:

\begin{equation}
R= \frac{k_BT}{6\pi \eta D} 
\label{eq2}
\end{equation}

where k$_B$ is the Boltzman constant, T is the sample temperature (T=295 K for these measurements) and $\eta$ is the solvent viscosity ($\eta$=0.9544 mPas at T=295 K \cite{Eta}). From this analysis a value of R$_{DLS}$=(36.0 $\pm$ 0.4) nm was obtained.

The SANS from non-interacting monodisperse spherical particles of uniform scattering length density is given by \cite{Guinier_Wiley_1955}:

\begin{equation}
I(Q)=\Phi V (\Delta\rho)^2 \left[\frac{3\: (sin(QR)-QR\:cos(QR))}{(QR)^3}\right] ^2 + bkgd
\label{eq3}
\end{equation}

where $\Phi$ is the volume fraction of particles, \textit{V} is the volume of one particle, $\Delta \rho$ is the difference in neutron scattering length density (SLD \cite{SLD}) between the particles ($\rho_{PSL}$=+1.42x10$^{10}$ cm$^{-2}$) and the dispersion medium ($\rho_{H2O}$=-0.56x10$^{10}$ cm$^{-2}$, $\rho_{D2O}$=+6.33x10$^{10}$ cm$^{-2}$), \textit{R} is the radius of the particles, and \textit{bkgd} is the residual $Q$-independent background. Where necessary Eq.\ref{eq3} can be integrated over a particle size distribution. Here polydispersity was not taken into account, assuming a highly monodisperse sample, but the estimated dQ data were used for instrumental resolution smearing.

Least-squares fitting Eq.\ref{eq3} to the SANS data returned a value of R$_{SANS}$=(34.5 $\pm$ 0.3) nm in very good agreement with the expected radius and with $R_{DLS}$. This gave confidence that the DLS-SANS apparatus was working as intended. Note that it is not uncommon for $R_{DLS}>R_{SANS}$ as the former typically includes a 'shell' of coordinated solvent molecules that moves with the particle and retards Brownian diffusion. The difference observed here is entirely consistent with measurements of the typical thickness of this solvent shell \cite{ThomaNC2019}.

\subsubsection{PNIPAM Microgel Sample}
\begin{figure}
\centering
\includegraphics[trim=0cm 20cm 0cm 0cm,clip, width=.99\textwidth]{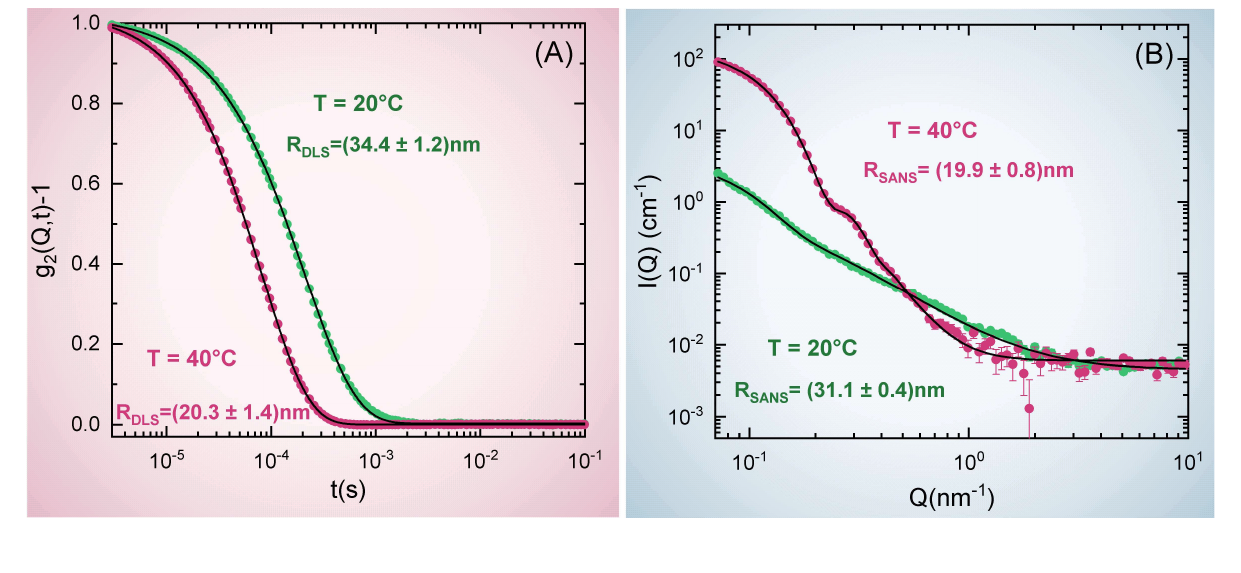}
\caption{(A) Intensity autocorrelation function and (B) SANS intensity as simultaneously obtained from the PNIPAM microgel sample at T=293 K (below the VPT) and at T=313 K (above the VPT) using the DLS-SANS apparatus. The continuous black lines superimposed on the data (symbols) are fits to Eq.\ref{eq1} and Eq.\ref{eq4}, respectively.} \label{f7}
\end{figure}

The second sample that has been investigated was an aqueous suspension of poly(N-isopropylacrylamide) (PNIPAM) microgel particles. PNIPAM microgels are thermoresponsive colloidal particles having rather different dimensions either side of a temperature-induced Volume Phase Transition (VPT). The particles collapse from a swollen to a shrunken state at T$\cong$305 K. This soft colloid, and its derivatives, has been widely studied in the literature from both a fundamental science perspective and for its many potentially innovative applications \cite{FernandezBook2011, RovigattiSM2018, BrijittaCOCIS2019}. For the present study a PNIPAM microgel was synthesized in our laboratory following the procedure fully described in Ref.\cite{NigroJCP2015, NigroCSA2017}. 
A dispersion of the microgel in D$_2$O at a weight concentration C$_w$=0.3\% was measured at T=293 K, below the VPT, and at T= 313 K, above the VPT. Representative examples of the resulting DLS intensity autocorrelation functions, $g_2(Q,t)-1$, are shown in Fig.~\ref{f7}(a) and of the SANS data, $I(Q)$, in Fig.~\ref{f7}(b). The SANS data were acquired over ~4 to 8 hours.  
The differences between the swollen and shrunken states are clearly evident in both cases. As before, the DLS data has been fitted to Eq.\ref{eq1} and Eq.\ref{eq2} to obtain the hydrodynamic sizes of the microgel particles. This yielded R$_{DLS}$=(34.4 $\pm$ 1.2) nm at T=293 K and R$_{DLS}$=(20.3 $\pm$ 1.4) nm at T=313 K. 

Interpreting these SANS data is, however, a little more complicated because, unlike the standard sample, the internal structure of the microgel particles is not homogeneous and the interface with the dispersion medium is less-defined. The former condition can be accounted for by assuming that the distance between cross-links in the microgel (the 'mesh size') can be described by a single density correlation length. The scattering from this is the well-known Ornstein-Zernicke model which has a Lorenztian form. The second condition can be accounted for by taking the scattering function for a sphere (the term in the square brackets in Eq.\ref{eq3}) and convoluting it with a decay function to 'soften' the drop-off in scattering length density at the periphery of the sphere. A common decay function that is applied is a Gaussian, and this leads to a model known as the "fuzzy sphere". The overall scattering function is then a sum of the two contributions \cite{PritiJCP2014, ScottiPNAS2016}:

\begin{equation}
I(Q) = \left[ \frac{\Phi}{V} \left[\frac{3 V \Delta\rho\: (sin(QR)-QR\:cos(QR))}{(QR)^3} exp \left(\frac{-(\sigma Q)^2}{2}\right) \right] ^2 + \left[\frac{I_L(0)}{1+(\xi Q)^2}\right]\right] + bkgd
\label{eq4}
\end{equation}

where $\Phi$, \textit{V}, and $\Delta \rho$ all have the same meaning as before (with the scattering length density of PNIPAM $\rho_{PNIPAM}$=+8.14x10$^{9}$ cm$^{-2}$). The parameter $\sigma$ is the 'fuzziness' of the particle interface; related to the distance over which the SLD decays from its starting value ($\sigma$ $\ll$ \textit{R}). The way that this is defined means that in this model \textit{R} is the radius at which the SLD profile has decreased by half. The overall radius of the microgel particle is then given by $R_{SANS}=R+2\sigma$ \cite{StiegerLangmuir2004}. The gel structure itself is characterised by the correlation length $\xi$. $I_{L}(0)$ is the intensity of the Lorentzian contribution at Q = 0. Where necessary Eq.\ref{eq4} can be integrated over a particle size distribution. Radius polydispersity (PD) was accounted for with a Schulz distribution and the estimated dQ data were once again used to account for instrumental resolution smearing.

Least-squares fitting Eq.\ref{eq4} to the SANS data returned values of R$_{SANS}$=(31.1 $\pm$ 0.4) nm with PD=0.25 at T=293 K and R$_{SANS}$=(19.9 $\pm$ 0.8) nm with PD=0.14 at T=313 K. These values are again in very good agreement with the hydrodynamic radii from the DLS. It is interesting to note that the agreement is better when the microgel particles are in the shrunken state, presumably because they then appear more homogeneous.

\section{CONCLUSION}
In this work we have presented a new Dynamic Light Scattering apparatus conceived to permit simultaneous DLS and SANS measurements, and demonstrated that it produces good quality and reliable data with test measurements on two different samples (a polystyrene latex calibration standard and a poly(N-isopropylacrylamide) microgel dispersion) on two SANS instruments at the ISIS Pulsed Neutron \& Muon Source. Using the DLS data to derive size information produced results in excellent agreement with the values obtained from the SANS data. However, the real benefit of DLS is its ability to provide information about the microscopic dynamical behaviour in soft matter samples, in particular in those which are sensitive to small changes in physicochemical stimuli. This will be the focus of future work. 

\section{AUTHOR'S CONTRIBUTIONS}
B.R., R.A. and S.K. conceived the idea and the scheme of the apparatus.
F.C. and R.S. provided the technical design and mechanically realized the apparatus that was assembled by F.B..
A.C. and T.C. provided the sample changer and technical assistance during the SANS experiments.
V.N. did the alignment.
V.N., R.A., E.B., S.F. and B.R. performed the measurements and the analysis with the help of S.K. and N.M..
B.R., V.N., S.K., R.A. and N.M. wrote the manuscript.
B.R. supervised the project.

\section{SUPPLEMENTARY MATERIAL}
Although commercial analysis software was provided with the correlator, as part of this project the KWW function (Eq.\ref{eq1}) and Cumulants analysis functions were generated as plugin models for the SasView program. They are free to download from http://marketplace.sasview.org/.

\section{ACKNOWLEDGEMENTS}
The authors would like to thank Iain Johnson, Maksim Schastny and Jamie Nutter (ISIS Sample Environment Electronics Team), and Alistair McGann, David Keymer, Thomas Willemsen and John Holt (ISIS Instrument Control Software Team) for their assistance.  
This work was supported within the CNR-STFC Agreement 2014–2020 (No. 3420)
concerning collaboration in scientific research at ISIS Spallation Neutron Source.
The authors also gratefully acknowledge
the UK Science \& Technology Facilities Council for the provision of neutron beam time at ISIS (Experiments RB1830612 and RB1920335).

\section{DATA AVAILABILITY}
The data that support the findings of this study are available from the corresponding author upon reasonable request.

%

\end{document}